\documentclass[reprint,amsmath,amssymb,pra, aps, showpacs,showkeys]{revtex4-1}

\usepackage{graphicx}
\usepackage{dcolumn}
\usepackage{bm}
\usepackage{xfrac}
\usepackage{tikz}
\usetikzlibrary{decorations.pathreplacing, shadows, shapes}
\newcommand{\ket}[1]{\left| #1 \right>} 
\newcommand{\braket}[2]{\left< #1 \vphantom{#2} \right| \left. #2 \vphantom{#1} \right>} 
\newcommand{\ketbra}[2]{|#1\rangle\langle#2|}

\usepackage{color}
\definecolor{myred}{RGB}{153, 0,0}

\begin{document}

\title{Prediction by linear regression on a quantum computer}

\author{Maria Schuld}
\email[E-mail me at: ]{schuld@ukzn.ac.za}
\thanks{The authors would like to thank Patrick Rebentrost for fruitful discussions.}
\affiliation{Quantum Research Group, School of Chemistry and Physics, University of KwaZulu-Natal, Durban 4000, South Africa}
\author{Ilya Sinayskiy} 
\affiliation{Quantum Research Group, School of Chemistry and Physics, University of KwaZulu-Natal, Durban 4000, South Africa}
\affiliation{National Institute for Theoretical Physics (NITheP), KwaZulu-Natal, 4001, South Africa}
\author{Francesco Petruccione}
\affiliation{Quantum Research Group, School of Chemistry and Physics, University of KwaZulu-Natal, Durban 4000, South Africa}
\affiliation{National Institute for Theoretical Physics (NITheP), KwaZulu-Natal, 4001, South Africa}

\begin{abstract}
We give an algorithm for prediction on a quantum computer which is based on a linear regression model with least squares optimisation. Opposed to related previous contributions suffering from the problem of reading out the optimal parameters of the fit, our scheme focuses on the machine learning task of guessing the output corresponding to a new input given examples of data points. Furthermore, we adapt the scheme to non-sparse data matrices that can be represented by low-rank approximations, and significantly improve the dependency on its condition number. The prediction result can be accessed through a single qubit measurement or used for further quantum information processing routines. The algorithm's runtime is logarithmic in the dimension of the input space as well as the size of the dataset size provided the data is given as quantum information as an input to the routine. 
\end{abstract}

\pacs{03.67.Ac, 07.05.Mh, 07.05.Pj}
\keywords{Quantum machine learning, pattern classification, image recognition, matrix inversion, quantum algorithm, linear regression, least squares}

\maketitle

\section{Introduction}
The central problem of machine learning is pattern recognition \cite{bishop06}, in which a machine is supposed to infer from a set of training data how to map new inputs of the same type to corresponding outputs. An important application lies in image recognition, where example images are presented to the computer together with a class label describing their content, and the task is to generalise from the training data in order to classify a new input image correctly. Nowadays, image recognition by machines can compete with human abilities for selected problems \footnote{See for example Yann LeCun's list of the performance of different machine learning methods on the MNIST handwritten digits database: yann.lecun.com/exdb/mnist.}. Pattern recognition plays an increasingly important role in IT and robotics, but also in fields such as medicine and finance - in short, everywhere a decision has to be derived from data. If the class label in a pattern recognition task is continuous, one speaks of a \textit{regression} problem on which we will focus here. \\
A widely used method adopted from statistics is linear regression, which predicts new labels based on the linear fit of the data. \textit{Learning} in this context refers to the process of estimating optimal parameters for the linear fit by reducing the least squares error of the data points. In computational terms, this reduces to finding the (pseudo)inverse of a matrix which represents the data, a task that can become very expensive considering the size of datasets in industrial applications today.\\
Quantum computing established itself as a promising extension to classical computation \cite{arora09}, and was theoretically shown to perform significantly better in selected computational problems. Recently, a number of proposals apply quantum information processing to methods in machine learning \cite{wiebe14, wiebe14b, ogorman15, lloyd13,schuld15a}. One line of approaches \cite{rebentrost14,  zhao15, wiebe12} focuses on the numerous machine learning methods formulated as matrix inversion problems, which can be tackled by the quantum algorithm to solve linear systems of equations introduced by Harrow, Hassidim and Lloyd (HHL) \cite{harrow09}. In particular, Wiebe, Braun and Lloyd (WBL) \cite{wiebe12} suggested a quantum algorithm for data fitting based on HHL. The authors approach linear regression from a statistics perspective, and their goal is to find a quantum state of which the amplitudes represent the optimal parameters of the linear fit learned from data. They show that this can be done in time logarithmic in the dimension of the input data $N$ (e.g., the number of pixels in image recognition) for a $s$ sparse data matrix given in the form of quantum information, and with a sensible dependency on its condition number $\kappa$ as well as the desired accuracy $\epsilon$ (roughly $\mathcal{O}(\log N s^3 \kappa^6\epsilon^{-1}  )$). They conclude that ``[a]lthough the algorithm yields [the state encoding the parameters] efficiently, it may be exponentially expensive to learn via tomography'' \cite{wiebe12} and shift their focus on a routine which estimates the quality of a fit. Another attempt in the same direction, but based on the singular value decomposition that we will also use here, is given in \cite{wang14} and requires a low rank data matrix instead of the sparseness condition. The author proposes various algorithms to analyse the parameter vector using known quantum routines. \\
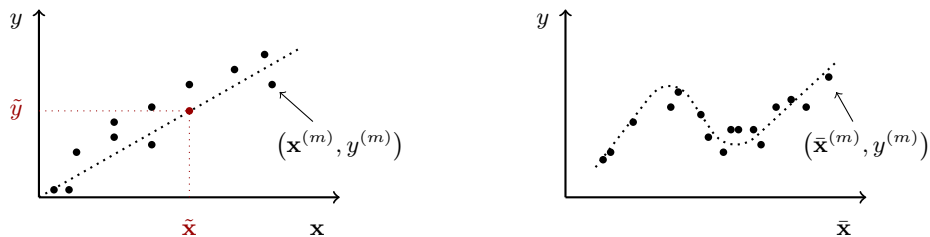
\begin{figure*}[t]
\centering
\begin{tikzpicture}
 \draw [->, thick](0,0)-- (4,0); 
 \draw [->,thick](0,0)-- (0,2.5); 
\path (-0.3,2.3) node[anchor=base] {$y$};
\path (3.7,-0.5) node[anchor=base] {$\mathbf{x}$};

\path (4,0.6) node[anchor=base](point) {$\big(\mathbf{x}^{(m)},y^{(m)}\big)$};
\draw[->] (point)--(3.2,1.4);

\draw[dotted, thick] (0,0) --(3.5,2);

\fill (2,1.5) circle (0.5mm);
\fill (1,1.0) circle (0.5mm);
\fill (0.4,0.1) circle (0.5mm);
\fill (1.5,0.7) circle (0.5mm);
\fill (1,0.8) circle (0.5mm);
\fill (0.2,0.1) circle (0.5mm);
\fill (0.5,0.6) circle (0.5mm);
\fill (1.5,1.2) circle (0.5mm);
\fill (2.6,1.7) circle (0.5mm);
\fill (3,1.9) circle (0.5mm);
\fill (3.1,1.5) circle (0.5mm);

\fill[myred] (2,1.15) circle (0.5mm);
\draw[dotted,myred] (2,0) --(2,1.15); 
\draw[dotted,myred] (0,1.15) --(2,1.15); 
\path (-0.3,1.1) node[anchor=base,myred] {$\tilde{y}$};
\path (2,-0.5) node[anchor=base,myred] {$\tilde{\mathbf{x}}$};

\draw [->, thick](7,0)-- (11,0); 
\draw [->,thick](7,0)-- (7,2.5); 
\path (6.7,2.3) node[anchor=base] {$y$};
\path (10.7,-0.5) node[anchor=base] {$\bar{\mathbf{x}}$};

\path (11,0.6) node[anchor=base](point) {$\big(\bar{\mathbf{x}}^{(m)}, y^{(m)}\big)$};
\draw[->] (point)--(10.6,1.4);

\draw[dotted, thick, rounded corners=13pt]  (7.4,0.4) --(8.4,1.7)--(9.2,0.5)--(10.6,1.8);

\fill (7.6,0.6) circle (0.5mm);
\fill (7.5,0.5) circle (0.5mm);
\fill (7.9,1.0) circle (0.5mm);
\fill (8.4,1.2) circle (0.5mm);
\fill (8.5,1.4) circle (0.5mm);
\fill (8.8,1.1) circle (0.5mm);
\fill (8.9,0.8) circle (0.5mm);
\fill (9.2,0.9) circle (0.5mm);
\fill (9.1,0.6) circle (0.5mm);
\fill (9.3,0.9) circle (0.5mm);
\fill (9.5,0.9) circle (0.5mm);
\fill (9.6,0.7) circle (0.5mm);
\fill (9.8,1.2) circle (0.5mm);
\fill (10.0,1.3) circle (0.5mm);
\fill (10.2,1.2) circle (0.5mm);
\fill (10.5,1.6) circle (0.5mm);

\end{tikzpicture}
\caption{Illustration of linear regression given training data points $(x^{(m)},y^{(m)})$ (left). Opposed to statistical analysis, the task in machine learning is not to determine the fit itself, but to guess the class label $\tilde{y}$ of a new input $\tilde{x}$ based on the fitted function. The inputs can be mapped from original inputs $\bar{\mathbf{x}}$ by a nonlinear function, dramatically increasing the power of linear regression to fit nonlinear functions (right).}
\label{Fig:reg}
\end{figure*}
In this Article, we want to add an important missing piece to the literature of quantum machine learning by reapproaching the rich problem of linear regression on a quantum computer from a machine learning perspective, in which the focus lies on the prediction of new inputs based on a dataset. We demonstrate a routine for prediction which is roughly in $\mathcal{O}( \log N \kappa^2 \epsilon^{-3})$, provided the information is given encoded into a quantum state. Using the techniques of quantum Principal Component Analysis proposed in \cite{lloyd14}, our algorithm does not require the data matrix $\mathbf{X}$ (defined below) to be sparse. Instead, we only need to be able to represent $\mathbf{X}^T\mathbf{X}$  by a low-rank approximation, meaning that it is dominated by a few large eigenvalues.    \\

\section{Prediction with linear regression}
The problem of supervised pattern recognition can be described as follows. Given a set of data points $\mathcal{D}=\{(\mathbf{x}^{(1)}, y^{(1)}),...,(\mathbf{x}^{(M)},y^{(M)})\}$ with $\mathbf{x}^{(m)} \in \mathbb{R}^N$ and $y^{(m)} \in \mathbb{R} $ for $m=1,...,M$, the goal is to predict the output $\tilde{y} \in \mathbb{R}$ to a new input $\tilde{\mathbf{x}} \in \mathbb{R}^N$. A linear model proposes a map 
\begin{equation}f(\mathbf{x}, \mathbf{w})= \mathbf{x}^T \mathbf{w} \label{Eq:lr}, \end{equation}
where $\mathbf{w} = (w_1, ...,w_N)^T\in \mathbb{R}^{N}$ is a vector of parameters that have to be `fitted' or learned from the data. (Note that we can omit a potential bias $w_0$ by including it in $\mathbf{w}$ and extending $\mathbf{x}$ by a corresponding entry $x_0=1$.) It is important to emphasise that the inputs $\mathbf{x}$ can be generated by a nonlinear map from some original space in which the data is given into $\mathbb{R}^N$, which enables linear regression to fit nonlinear functions (e.g. in polynomial curve fitting, see Figure \ref{Fig:reg}). \\
The most common method  to estimate the parameters (yielding the minimum variance with zero bias for the estimate) is to minimise the least squares error between the predicted values by the model, $f(\mathbf{x}^{(m)}, \mathbf{w})$, and the actual target outputs $y^{(m)}$ for each datapoint,
\[\min_{\mathbf{w}} \; \sum\limits_{m=1}^M (f(\mathbf{x}^{(m)}, \mathbf{w}) - y^{(m)})^2.\] As apparent from this objective function we focus on un-regularised linear regression here, and the matter of regularisation is an open question in quantum machine learning research. \footnote{In machine learning applications, a regularisation term favouring small parameters is usually added to avoid overfitting. It is an interesting question for further research whether the normalisation condition for quantum states (i.e., containing the parameters as in WBL) can be harvested to introduce an effective regularisation.}\\
Introducing the compact matrix notation $\mathbf{X} = (\mathbf{x}^{(1)},..., \mathbf{x}^{(M)})^T$ (which we refer to as the data matrix containing all training inputs as rows) and the target vector $\mathbf{y}= (y^{(1)},...,y^{(M)})^T$, the least squares error becomes $|\mathbf{X}\mathbf{w} - \mathbf{y}|^2$, and it is a widely used result \cite{golub70} that the solution to the least squares problem is given by
\[\mathbf{w} = \mathbf{X}^+ \mathbf{y}, \]
where $ \mathbf{X}^+$ is the Moore-Penrose pseudoinverse of $\mathbf{X}$. Formally, one can write $\mathbf{X}^+ = (\mathbf{X}^{\dagger}\mathbf{X})^{-1} \mathbf{X}^{\dagger}$ if $(\mathbf{X}^{\dagger}\mathbf{X})^{-1}$ exists, and the approach in WBL solves the resulting equation via quantum computation. We will use a different form of the Moore-Penrose pseudoinverse based on the reduced singular value decomposition $\mathbf{X} = \mathbf{U} \mathbf{\Sigma} \mathbf{V}^{\dagger}$ \cite{hogben06} in order to allow for $\mathbf{X}$ to be non-sparse. Here, $\mathbf{\Sigma}$ is a diagonal matrix containing the real singular values $\sigma_1, \sigma_2,... ,\sigma_R>0$ and the $r$th orthogonal column of $\mathbf{U} \in  \mathbb{R}^{M\times R}$ [$\mathbf{V} \in  \mathbb{R}^{J\times R}$] is the $r$th left [right] eigenvector $ \mathbf{v}_r$  [$\mathbf{u}_r$] to the singular value $\sigma_r$. As opposed to the eigenvalue decomposition, the singular value decomposition can always be found, and in particular for non-square matrices \cite{trefethen97}. The pseudoinverse can then be defined as $ \mathbf{X}^+ =  \mathbf{V} \mathbf{\Sigma}^{-1} \mathbf{U}^{\dagger}$. The singular values of  $\mathbf{X}$ are at the same time the square roots of the nonzero eigenvalues  $\{\sqrt{\lambda_1},  ..., \sqrt{\lambda_R}\}$ of $\mathbf{X}^{\dagger}\mathbf{X}$ and $\mathbf{X}\mathbf{X}^{\dagger}$, and the left and right singular vectors are their respective eigenvectors; a fact of which we will make use below. \\
Using this mathematical background in the case of real spaces, as well as the convenient alternative formulation $\mathbf{X}= \sum_r \sigma_r \mathbf{u}_r \mathbf{v}_r^{T}$ \cite{trefethen97}, the least squares solution can be written as
\begin{equation}
\mathbf{w} = \sum\limits_{r=1}^{R} \sigma_r^{-1}  \mathbf{v}_r \mathbf{u}_r^{T} \mathbf{y}.
\label{Eq:w}
\end{equation}
Moreover, according to Eq. (\ref{Eq:lr}) the output $\tilde{y}$ for a new input $\tilde{\mathbf{x}}$ is given by
\begin{equation}
\tilde{y} = \sum\limits_{r=1}^{R} \sigma_r^{-1} \, \tilde{\mathbf{x}}^T \mathbf{v}_r \mathbf{u}_r^{T} \mathbf{y}.
\label{Eq:ynew}
\end{equation}
The desired output (`prediction') of the pattern classification algorithm is thus a single scalar value, which will circumvent the problem of expensive state tomography notorious for quantum machine learning algorithms \cite{aaronson15}. The problem stated here can be solved in polynomial time $\mathcal{O}(N^{d})$ on a classical computer, where for the best current algorithms $2 \leq d \leq 3$. The aim of this work is to find a quantum algorithm that can reproduce the result from Eq. (\ref{Eq:ynew}) more efficiently assuming that the data is given as quantum information in the form specified below. 

\section{The quantum linear regression algorithm} 
The quantum linear regression algorithm is based on the method of encoding classical information such as a $2^n$ dimensional vector $\mathbf{a} =(a_0,...,a_{2^n-1})^T$ into the $2^n$ amplitudes $a_0,...,a_{2^n-1}$ of a $n$-qubit quantum system, $\ket{\psi_{\mathbf{a}}} = \sum_{i=0}^{2^n-1} a_i \ket{i}$, where $\{\ket{i}\}$ is a convenient notation for the computational basis $\{\ket{0...0} \hat{=} \ket{0},...,\ket{1...1} \hat{=} \ket{2^n-1}\}$. In other words, the probabilistic description of a set of 2-level quantum systems is used to store and manipulate classical information. We will refer to this method here as \textit{amplitude encoding} and denote every such quantum state by a $\psi$ with a subscript referring to the classical vector it encodes. (In other words, $\ket{a}$ is a quantum state where some mathematical object $a$ is encoded into the basis state in a way that is to specify in detail, while $\ket{\psi_{\mathbf{a}}}$ is a quantum state representing the real vector $\mathbf{a}$ via amplitude encoding.)  \\
The general idea of the quantum pattern recognition algorithm is to create a quantum state representing the data matrix $\mathbf{X}= \sum_r \sigma_r \mathbf{u}_r \mathbf{v}_r^{T}$ via amplitude encoding (Step A) and use tricks from \cite{lloyd14,harrow09} to invert the unknown singular values efficiently (Step B and C). In Step D, quantum state representations of $\mathbf{y},\tilde{\mathbf{x}}$ are used to write the desired prediction from Eq. (\ref{Eq:ynew}) into the offdiagonal elements of an ancilla qubit, where it can be read out by a simple $\sigma_x, \sigma_y$ measurement. Details regarding the computational complexity will be considered further below.

\subsection{State preparation}  
The quantum algorithm takes copies of the quantum states representing each of the objects $\mathbf{X}, \mathbf{y}$  and   $\tilde{\mathbf{x}}$ from above in amplitude encoding,
\begin{align}\ket{\psi_{\mathbf{X}}} &=\sum\limits_{j=0}^{N-1}  \sum\limits_{m=0}^{M-1} x_j^{(m)} \ket{j} \ket{m}, \label{Eq:psiX}\\
\ket{\psi_{\mathbf{y}}} &= \sum\limits_{\mu=0}^{M-1} y^{(\mu)} \ket{\mu},\label{Eq:psiy}\\
\ket{\psi_{\tilde{\mathbf{x}}}} &= \sum\limits_{\gamma=0}^{N-1} \tilde{x}_{\gamma} \ket{\gamma},\label{Eq:psix}
\end{align}
with $\sum_{m,j} | x_j^{(m)} |^2=\sum_{\mu} |y^{(\mu)}|^2=\sum_{\gamma} | \tilde{x}_{\gamma}|^2=1$. Note that the algorithm thus works with normalised data, and the results have to be re-scaled accordingly. Using the Gram-Schmidt decomposition, we can formally write 
\[\ket{\psi_{\mathbf{X}}}  =\sum_{r=1}^{R} \sigma^r \sum\limits_{j=1}^{J}v^r_j\ket{j}\sum\limits_{m=1}^{M}u^r_m \ket{m},\] 
or in short, 
\begin{equation}
\ket{\psi_{\mathbf{X}}}= \sum_{r=1}^{R} \sigma^r \ket{\psi_{\mathbf{v}^r} }  \ket{\psi_{\mathbf{u}^r} } . \label{Eq:gs}
\end{equation} 
Here, $\ket{\psi_{\mathbf{v}^r} }  =\sum_{j=1}^{J} v^r_j\ket{j}$ and $\ket{\psi_{\mathbf{u}^r} } =\sum_{m=1}^{M}u^r_m \ket{m}$ are quantum states representing the orthogonal sets of left and right singular vectors of $\mathbf{X}$ (and at the same time the eigenvectors of $\mathbf{X}^{\dagger}\mathbf{X}$ as well as $\mathbf{X}\mathbf{X}^{\dagger}$) via amplitude encoding, and $\sigma^r$ are the corresponding singular values. The number of qubits needed to construct states (\ref{Eq:psiX}-\ref{Eq:psix}) are $\lceil\log N \rceil+ \lceil\log M\rceil$, $\lceil\log M\rceil$ and $\lceil\log N \rceil$ respectively, and the aim of the following algorithm is to remain linear in the number of qubits or logarithmic in the problem size. The states (\ref{Eq:psiX}-\ref{Eq:psix}) can be understood as the result of previous quantum computation or simulation as argued in \cite{wiebe12}. If given a `classical' dataset, techniques for the efficient preparation of arbitrary initial quantum states are a nontrivial and still controversially discussed topic, although some ideas have been brought forward of how to prepare certain states linear in the number of  qubits \cite{grover02} or via quantum Random Access Memories \cite{giovannetti08}. Clader et al \cite{clader13} show a way to prepare a quantum state efficiently for the HHL algorithm if an oracle is given that conditioned on a register in uniform superposition `loads' the vector entries into an entangled register, `writes' these entries into the amplitude of an ancilla and performs a conditional measurement to prepare the desired state in amplitude encoding. This conditional measurement has a high probability to succeed for relatively uniform entries, while sparse amplitudes underlie the Grover bound and cannot be prepared in time logarithmic in the dimension of the classical vector we want to encode into a quantum state.  We acknowledge that overall, the question of state preparation is an outstanding challenge for quantum machine learning algorithmic design.

\subsection{Extracting the singular values} 
In order to transform Eq. (\ref{Eq:gs}) into a `quantum representation' of the result (\ref{Eq:ynew}) we need to first invert the singular values of $\mathbf{X}$. For this, we will `extract' the eigenvalues $\lambda_r$ of $\mathbf{X}^{\dagger}\mathbf{X}$ to eigenvectors $\mathbf{v}_r$ and use the inversion procedure from \cite{harrow09} in the following step. In order to access the eigenvalues, we use copies of the state (\ref{Eq:psiX}) in which on the level of description we ignore the $\ket{m}$ register in order to obtain a mixed state $\rho_{\mathbf{X}^{\dagger}\mathbf{X}} = \mathrm{tr}_m \{ \ketbra{\psi_{\mathbf{X}}}{\psi_{\mathbf{X}}}\} $ which represents the positive hermitian matrix $\mathbf{X}^{\dagger}\mathbf{X}$,
\[ \rho_{\mathbf{X}^{\dagger}\mathbf{X}} = \sum\limits_{j,j'=1}^{N} \sum\limits_{m=1}^M x_j^{(m)} x_{j'}^{(m)*} \ketbra{j}{j'}.\] 
From here, we use the ideas of \textit{quantum Principal Component Analysis} introduced in \cite{lloyd14} to `apply' $\rho_{\mathbf{X}^{\dagger}\mathbf{X}}$ to $\ket{\psi_{\mathbf{X}}}$, resulting in
\[ \sum \limits_{k=0}^{K} \ketbra{k\Delta t}{k\Delta t} \otimes e^{-i k\rho_{\mathbf{X}^{\dagger}\mathbf{X}}\Delta t}  \ketbra{\psi_{\mathbf{X}}}{\psi_{\mathbf{X}}}  e^{i k\rho_{\mathbf{X}^{\dagger}\mathbf{X}}\Delta t}, \] for some large $K$. As outlined in \cite{lloyd14}, the quantum phase estimation algorithm results in
\[\sum_{r=1}^R \sigma^r \ket{\psi_{\mathbf{v}^r} }   \ket{\psi_{\mathbf{u}^r} }  \ket{\lambda^r} , \]
in which the eigenvalues $\lambda^r = (\sigma^r)^2$ of $\rho_{\mathbf{X}^{\dagger}\mathbf{X}} $ are approximately encoded in the $\tau$ qubits of an extra third register that was initially in the ground state.

\subsection{Inverting the singular values} 
Adding an extra qubit and rotating it conditional on the eigenvalue register \cite{harrow09} then yields
\begin{equation} \sum_{r=1}^R \sigma^r \ket{\psi_{\mathbf{v}^r} }  \ket{\psi_{\mathbf{u}^r} }  \ket{\lambda^r} \left( \sqrt{1- \left(\frac{c}{\lambda^r}\right)^2}\ket{0} +\frac{c}{\lambda^r} \ket{1}\right) .\end{equation}
The constant $c$ is chosen so that the inverse eigenvalues are not larger than $1$, which is given if it is smaller than the smallest nonzero eigenvalue $\lambda^{\mathrm{min}}$ of $\mathbf{X}^{\dagger}\mathbf{X}$, or equivalently, the smallest nonzero squared singular value $(\sigma^{\mathrm{min}})^2$ of $\mathbf{X}$. We perform a conditional measurement on the ancilla qubit, only continuing the algorithm (`accepting') if the ancilla is in state $\ket{1}$ (else the entire procedure has to be repeated). Note that amplitude amplification can boost the probability of accepting quadratically as discussed in the runtime analysis below. Uncomputing and discarding the eigenvalue register results in
\begin{equation} \ket{\psi_1} := \frac{1}{\sqrt{p(1)}}\sum_{r=1}^R \frac{c}{\sigma^r} \ket{\psi_{\mathbf{v}^r} }  \ket{\psi_{\mathbf{u}^r} } ,\label{Eq:res3}\end{equation}
where the probability of acceptance is given by $p(1)= \sum\limits_r \left|\frac{c}{\lambda^r}\right|^2 $.\\

\subsection{Executing the inner products} 
The last step has the goal to write the desired result $\sum_{r=1}^R (\sigma^r)^{-1} \braket{\psi_{\tilde{\mathbf{x}}}}{v^r}   \braket{\psi_{\mathbf{y}}}{u^r}$ into selected entries of an ancilla's single qubit density matrix, from which it can be accessed by a simple measurement. Consider the result (\ref{Eq:res3}) of the previous step, as well as $\ket{\psi_2} =\ket{\psi_{\mathbf{y}}}\ket{\psi_{\tilde{\mathbf{x}}}} $ from (\ref{Eq:psiy},\ref{Eq:psix}).  The inner product cannot be extracted by a direct SWAP test between $\ket{\psi_1}$ and $\ket{\psi_2}$ if we want the output domain to extend into the negative numbers, since this would lead to an acceptance probability of $\frac{1}{2} + \frac{1}{2} |\braket{\psi_1}{\psi_2}|^2$, which is ambiguous in the sign of the desired result $\braket{\psi_1}{\psi_2}$. The follwing `prediction routine' circumvents these problems: Conditionally prepare the two states so that they are entangled with an ancilla qubit, 
\begin{equation} \sfrac{1}{\sqrt{2}}( \ket{\psi_1} \ket{0} + \ket{\psi_2} \ket{1}), \label{Eq:ent}
\end{equation}
and trace out all registers except from the ancilla, the offdiagonal elements $\rho_{12}, \rho_{21}$ of the ancilla's density matrix read
\[ \frac{c}{2\sqrt{p(1)}}\sum_{r} (\sigma^r)^{-1}   \sum_{j}  v^r_{j}\tilde{x}_{j}     \sum_{m} u^r_{m}  y^{(m')},  \]
and contain the desired result $(\ref{Eq:ynew})$ up to a known normalisation factor. Conditionally preparing (\ref{Eq:ent}) requires us to execute the entire algorithm including state preparation conditioned on the state of the ancilla qubit and might not be easy to implement. In that case one can adapt the algorithm so that the $\ket{0...0}$ basis state in $\ket{\psi_1}$ and $\ket{\psi_2}$ is `excluded' from all operations and remains with a constant amplitude $\frac{1}{\sqrt{2}}$ throughout the algorithm (while the other $2^n-1$ amplitudes are renormalised accordingly). This prepares states of the general form $\ket{a} = \frac{1}{\sqrt{2}}(\ket{0...0} + \sum_{i=1}^N a_i \ket{i}), \ket{b} =\frac{1}{\sqrt{2}}(\ket{0...0} +\sum_{j=0}^N b_j \ket{j})$. A common swap test effectively shifts the inner product by $1/2$ and thus reveals $|\braket{a}{b}|^2 = |\frac{1}{2} + \frac{1}{2}\sum_{i=1}^N a_ib_i|^2$ from which the sign of $\sum_{i=1}^N a_ib_i$ can be extracted.

\section{Runtime analysis}
According to \cite{lloyd14} one needs temporal resources $t = k \Delta t$ in $\mathcal{O}(\log N)$ and of the order $\mathcal{O}(\epsilon^{-3})$ copies of $\rho_{\mathbf{X}^{\dagger}\mathbf{X}}$ to `exponentiate' a density matrix in Step 2, where $\epsilon$ is the error and $N$ the dimension of the inputs in our data set. The method requires the density matrix $ \rho_{\mathbf{X}^{\dagger}\mathbf{X}}$ to be close to a low-rank approximation which is dominated by a few large eigenvalues in order to maintain the exponential speedup. In general, it takes time $t = \mathcal{O}(1/\delta)$ to simulate $e^{iHt}$ for a Hamiltonian $H$ up to error $\delta$ \cite{harrow09}, and with the trick from \cite{lloyd14} it takes time $t^2$ to do the same for $e^{i\rho t}$. This means that if we want to resolve relatively uniform eigenvalues of the order of $1/N$, time grows quadratically with $N$ and the exponential speedup is lost. Hence, the method is only efficient if the density matrix is dominated by a few large eigenvalues.\\ 
The singular value inversion procedure in Step 3 determines the runtime's dependency on the condition number of $\mathbf{X}$, $\kappa = \sigma^{\mathrm{max}}(\sigma^{\mathrm{min}})^{-1}$. The probability to measure the ancilla in the excited state is
\[p(1) = \sum\limits_r \left|\frac{c}{\lambda^r}\right|^2 \leq R \left|\frac{\lambda^{\mathrm{min}}}{\lambda^{\mathrm{max}}}\right|^2  =  R \kappa^{-4}, \] 
which means one needs on average less than $\kappa^4$ tries to accept the conditional measurement. Amplitude amplification as in \cite{harrow09,rebentrost14} reduces this to a factor of $\mathcal{O}(\kappa^2)$ in the runtime, which can become significant for matrices that are close to being singular. The condition number is a measure for how easy it is to invert a matrix, and it is no surprise that the runtime of the quantum algorithm also depends on it. The amended SWAP routine in Step D is also linear in the number of qubits, and the final measurement only accounts for a constant factor. The upper bound for the runtime can thus be roughly estimated as $\mathcal{O}(\log N \kappa^2 \epsilon^{-3})$ if we have sufficient copies of $\rho_{\mathbf{X}^{\dagger}\mathbf{X}}$ available which is required to be close to a low-rank matrix. Remember that this does not include the costs of quantum state preparation in case the algorithm processes classical information. Compared to the previous result in WBL, this is an improvement of a factor $\kappa^{-4}$ whereas the dependence on the accuracy is worse by a factor $\epsilon^{-2}$. However, opposed to other proposals our algorithm tackles the problem of pattern recognition or prediction. Furthermore, we can apply our algorithm efficiently to non-sparse, but low rank approximations of the matrix $\mathbf{X}^{\dagger}\mathbf{X}$. \\

\section{Conclusion}
We described an algorithm for a universal quantum computer to implement a linear regression model for supervised pattern recognition. The quantum algorithm reproduces the prediction result of a classical linear regression method with (unregularised) least squares optimisation, thereby covering an important area of machine learning. It runs in time logarithmic in the dimension $N$ of the feature vectors as well as independent of the size of the training set if the inputs are given as quantum information. Instead of requiring the matrix containing the training inputs, $\mathbf{X}$, to be sparse it merely needs $\mathbf{X}^{\dagger} \mathbf{X}$ to be representable by a low-rank approximation. One can furthermore transform the input data by a nonlinear feature map known as the ``kernel trick'' (discussed in \cite{rebentrost14} for polynomial kernels) to increase the potential power of the method. The application of different kernels as well as the question of how to include regularisation terms is still open for further research. The sensitive dependency on the accuracy as well as the unresolved problem of state preparation (which appears in any of the numerous quantum algorithms encoding classical information into the amplitudes of quantum states), illustrate how careful one needs to treat `magic' exponential speedups for pattern recognition. However, as demonstrated here, quantum information can make a contribution to certain problems of machine learning, promising further fruitful results in the emerging discipline of quantum machine learning.


\begin{acknowledgments}
This work is based upon research supported by the South African Research Chair Initiative of the Department of Science and Technology and National Research Foundation, as well as the European Commission.
\end{acknowledgments}

\end{document}